\documentstyle[12pt]{article}
\title{\baselineskip=9mm
Probing anharmonic properties of nuclear surface vibration 
by heavy-ion fusion reactions} 
\author{N.  Takigawa$^{1}$, K.  Hagino$^{1}$, and S.  Kuyucak$^{2}$ \\ \\
\medskip
{\it $^{1}$Department of Physics,
Tohoku University, Sendai 980--77, Japan}
\\
{\it $^{2}$Department of Theoretical Physics, Research School of Physical
Sciences, } \\
{\it Australian National University, Canberra,
ACT 0200, Australia}
}
\date{}

\begin{document}

\maketitle

\begin{center}
{\bf Abstract}
\end{center}

Describing fusion reactions between $^{16}$O and $^{154}$Dy 
and, between $^{16}$O and $^{144}$Sm by the $sd-$ and $sdf-$ 
interacting boson model, 
we show that heavy-ion fusion reactions are strongly affected by 
anharmonic properties of nuclear surface vibrations and nuclear shape,
and thus provide a powerful method to study details of nuclear structure 
and dynamics. 


\medskip

\begin{flushleft} 
{\bf 1. Introduction}
\end{flushleft}

Extensive experimental as well as theoretical studies of 
heavy-ion fusion reactions at energies near and below the Coulomb 
barrier in the past decade have offered many basic ideas on 
the effects of nuclear collective excitations such as surface vibrations 
and rotation, or particle transfer on the fusion cross section\cite{B88}. 
Recent very accurate data at ANU and at Legnaro are 
providing us with the possibility to step up to a more advanced stage of the 
research. For example, one can extract detailed informations on 
nuclear structure and dynamics from the data of heavy-ion fusion reactions. 
For instance, the sign of the hexadecupole deformation can be determined 
by studying the fusion barrier distributon, as was exemplified for the 
$^{16}$O + $^{186}$W, $^{154}$Sm fusion reactions \cite{LLW93,LDH95}. 
Also, Stefanini et al. \cite{SACN95} demonstrated 
the important role played by the double phonon excitations 
in the $^{58}$Ni + $^{60}$Ni fusion reactions.  
In this contribution we address to the effects of 
anharmonic properties of nuclear surface vibrations. Our study 
was motivated by the discussions with Jack, Nand and David. 
When we visited Canberra last spring, they introduced 
us their puzzle about the role of octupole vibrations in 
the fusion reactions between  $^{16}$O and $^{208}$Pb 
and between $^{16}$O and $^{144}$Sm. 
The problem is the following: in the former, a better agreement 
between data and theory concerning the fusion barrier distribution 
is obtained if one includes the double phonon coupling\cite{D97}. 
On the other hand, in the latter, a good agreement obtained 
by a single phonon coupling is destroyed if one adds the double 
phonon coupling \cite{M95}. $^{144}$Sm has low lying quartet states 
which seem to correspond to the double octupole phonon 
excitations\cite{GVB90}. 
Therefore, it is quite puzzling that the inclusion of the double phonon 
coupling spoils the agreement bewteen the data and theory. 
However, there exists a noticeable anharmonicity 
in the energy spectrum\cite{GVB90}. We therefore expected that 
the puzzle raised by Jack {\it et al.} can be attributed to 
the anharmonic properties of nuclear surface vibrations. 
This is what we are going to report in this contribution. Actually 
important anharmonic properties are concerned not with the energy 
spectrum, but rather with the transition matrices, especially 
reorientation effects. 

\medskip

\begin{flushleft} 
{\bf 2. Coupled-channels formalism in the IBM -- a case study for 
$^{16}$O + $^{154}$Dy fusion reactions}
\end{flushleft}

In this section we make a case study for the $^{16}$O + $^{154}$Dy fusion 
reactions by using the U(5) limit of the $sd-$ IBM. It is known that the 
U(5) limit of the IBM corresponds to the anharmonic vibrator 
in the geometrical model of Bohr and Mottelson \cite{CW88}. We choose the IBM 
to reduce the number of free parameters. 

We assume the following Hamiltonian for the total system 
\begin{equation}
H=-\frac{\hbar^2}{2\mu}\nabla^2+H_{IBM}+
V_{coup}({\mbox{\boldmath $r$}},\xi),
\label{htotalsd}
\end{equation}
where ${\mbox{\boldmath $r$}}$ is the coordinate of the relative motion 
between the projectile and target, $\mu$ is the reduced mass, and $\xi$ 
represents the internal degrees of freedom of the target nucleus.  $H_{IBM}$ 
is the IBM Hamiltonian for the quadrupole vibration in the 
target nucleus, for which we assume the harmonic limit
\begin{equation}
H_{IBM}=\epsilon_d\hat{n}_d 
\label{hibm}
\end{equation}
Here $\hat{n}_d$ is the number operator of the $d$ bosons and 
$\epsilon_d$ is the excitation energy of the quadrupole 
vibration.  
In the numerical analyses we also introduce a 
deviation from the harmonic spectrum and study its effects. 

The coupling between the relative motion and the intrinsic motion of the 
target nucleus is described by $V_{coup}$ in Eq. (\ref{htotalsd}). 
We use the linear coupling model and the no-Coriolis approximation 
\cite{TI86}. We then have 
\begin{equation}
V_{coup}({\mbox{\boldmath $r$}},\xi)=
U_N(r)+\frac{Z_PZ_Te^2}{r}+\frac{\beta}{\sqrt{4\pi N}}f(r)\hat{Q}_{20}
\label{vcoup}
\end{equation}
where the coupling form factor $f(r)$ consists of the nuclear and 
Coulomb parts and read 
\begin{equation}
f(r)=-R_T\frac{dU_N}{dr}+\frac{3}{5}Z_PZ_Te^2\frac{R_T^2}{r^3}
\label{cf}
\end{equation}
$N$ is the boson number, the subscripts $P$ and $T$ refer to the 
projectile and target nuclei, respectively.  The scaling of the coupling 
strength with $\sqrt{N}$ is introduced to ensure 
the equivalence of the IBM and the geometric model results in the large $N$ 
limit \cite{BBK94}.  $\beta$ is the quadrupole deformation parameter. 
We represent the quadrupole moment operator $\hat{Q}_{2\mu}$ by 
\begin{equation}
\hat{Q}_{2\mu}=s^{\dagger}\tilde{d}_{\mu} + sd^{\dagger}_{\mu} +
\chi(d^{\dagger}\tilde{d})^{(2\mu)}
\label{qop}
\end{equation}
where tilde is defined as $\tilde{b}_{l\mu}=(-)^{l+\mu}b_{l-\mu}$.  When 
the $\chi$ parameter is zero, the quadrupole moments of 
all states vanish, and one obtains the harmonic limit 
in the large $N$ limit. Non-zero values of the $\chi$ generate quadrupole 
moments and are responsible for the anharmonicities.
This can be clearly seen from the structure of the matrix elements 
appearing in the coupled-channels calculations, which read, for example, 
\begin{equation}
H_{IBM}+V_{coup}=
U_N(r)+\frac{Z_PZ_Te^2}{r}+
\left(\begin{array}{ccc}
0&F(r)&0\\
F(r)&\epsilon_d-\frac{2}{\sqrt{14N}}\chi F(r)
&\sqrt{2(1-1/N)}F(r)\\
0&\sqrt{2(1-1/N)}F(r)&2\epsilon_d+\delta-\frac{4}{\sqrt{14N}}\chi F(r)
\end{array}\right)
\label{hmat}
\end{equation}
if we truncate by the two phonon states. 
$F(r)$ is $\frac{\beta}{\sqrt{4\pi}}f(r)$. 
We introduced a parameter 
$\delta $ in order to represent the deviation of the energy spectrum from the 
harmonic limit. 
  
We now apply this formalism to the fusion reactions between 
$^{16}$O and $^{154}$Dy treating $^{16}$O as inert\cite{HTDHL97b}. The 
boson number for $^{154}$Dy is estimated to be 4 by considering 
64 and 82 as the proton and the neutron magic numbers, respectively. 
The data of the E2 transition and the intrinsic quadrupole moment 
lead to $\beta$=0.24 and $\chi$=-2.36. The excitation energy of the 
$d$ boson, $\epsilon_d$, has been identified with the excitation 
energy of the first excited state of $^{154}$Dy, 0.335 MeV. 

Fig.1 shows the role of anharmonicity 
in this reaction by comparing the fusion excitation function 
and the fusion barrier distribution calculated in four different ways. 
The dotted line is the result of the one phonon coupling 
in the harmonic oscillator limit. 
The dott-dashed line is the result of the one phonon coupling where 
the reorientation term is included. 
The dashed line is the result when one takes 
double phonen states into account in the harmonic oscillator limit. 
The solid line is the corresponding calculations where the 
anharmonicities are taken into account. 
The effects of anharmonicity can be hardly seen 
in the fusion excitation function. However, the fusion barrier distribution 
clearly shows a significant change from the harmonic oscillator limit 
by the anharmonicity.

Fig.2 shows the dependence of the fusion barrier distribution on the 
deviation of the energy spectrum from the harmonic oscillator limit 
(the upper panel) and on the boson number (the lower panel).  
The upper panel compares the fusion barrier distribution calculated for 
four different choices of the parameter $\delta$ in Eq. (\ref{hmat}). 
It shows that 
the fusion barrier distribution is insensitive to the anharmonicity 
concerning the energy spectrum.  
The lower panel compares the fusion barrier distribution calculated 
for various boson numbers $N$. For comparison, it also 
contains the fusion barrier distribution calculated in the harmonic 
oscillator limit in the case, where there exist one and two phonon states. 
One clearly observes a strong dependence of the fusion barrier distribution 
on the boson number, and also the effects of anharmonicity.  

Fig.3 shows how the sign of nuclear deformation, {\it i.e.} the sign of the 
parameter $\chi$ in the quadrupole moment (see Eq. (\ref{qop})), affects 
the excitation function of the fusion cross section and the fusion 
barrier distribution. Though the former is almost insensitive to the 
sign of the deformation, the latter drastically changes with 
the sign of the deformation. This enables us to determine 
the shape of nuclei with anharmonic vibrational modes of excitation 
by analysing the fusion barrier distribution. 

\medskip

\begin{flushleft} 
{\bf 3. Anharmonic effects of $^{144}$Sm on the $^{16}$O + $^{144}$Sm fusion 
reactions} 
\end{flushleft} 

We now come to the question raised by Jack {\it et al.} 
(see Ref.\cite{HTK97} for details). Our main concern is the 
anharmonic effects of the octupole excitation of $^{144}$Sm. We therefore 
extend the model in the previous section by including the $f$ bosons, and 
assume that the IBM Hamiltonian is given by 
\begin{equation}
H_{IBM}=\epsilon_d\hat{n}_d + \epsilon_f\hat{n}_f.
\label{hibmsdf}
\end{equation}
where $\hat{n}_f$ is the number operator for $f$ bosons 
and $\epsilon_f$ is the excitation energy of 
the octupole vibration.  Since the study in the previous section 
has shown that the anharmonicity concerning the energy spectrum 
does not so much influence both the excitation function of the 
fusion cross section 
and the fusion  barrier distribution, we assume here a harmonic spectrum. 
The coupling Hamiltonian now includes also the coupling to the octupole 
vibration. Furher, we modify the linear coupling model 
in the previous section 
by treating the nuclear coupling to full order, 
while we keep the Coulomb coupling in the linear coupling 
approximation\cite{BBK94,HTDHL97}.
The Coulomb and the nuclear parts of the coupling Hamiltonian 
are now given by 
\begin{eqnarray}
&&V_C(r,\xi)=\frac{Z_PZ_Te^2}{r}\left(1
+ \frac{3}{5}\frac{R_T^2}{r^2} \frac{\beta_2 \hat{Q}_{20}}{\sqrt{4\pi N}}
+ \frac{3}{7}\frac{R_T^3}{r^3} \frac{\beta_3 \hat{Q}_{30}}{\sqrt{4\pi N}} 
\right) \label{vc}, \nonumber \\
&&V_N(r,\xi)=-V_0 \left[ 1+\exp \left(\frac{1}{a} \left( r-R_0 - R_T ( 
\beta_2 \hat{Q}_{20} + \beta_3 \hat{Q}_{30})/ \sqrt{4\pi N} \right) 
\right)\right]^{-1}.
\label{v}
\end{eqnarray}
$\beta_2$ and $\beta_3$ in Eq. (\ref{v}) are the 
quadrupole and octupole deformation parameters, which are usually estimated 
from the electric transition probabilities using the expression 
$\beta_{\lambda}=4\pi(B(E\lambda)\uparrow)^{1/2}/3Z_T e R_T^\lambda$.  
However, this formula does not hold for anharmonic vibrators. Therefore, we 
treat $\beta_2$ and $\beta_3$ as free parameters and look for their optimal 
values to reproduce the experimental data.  $\hat{Q}_2$ and 
$\hat{Q}_3$ in Eq. (\ref{v}) are the quadrupole and the octupole operators in 
the IBM, which we take as
\begin{eqnarray}
&&\hat{Q}_2=s^{\dagger}\tilde{d} + sd^{\dagger} +
\chi_2(d^{\dagger}\tilde{d})^{(2)}
+ \chi_{2f}(f^{\dagger}\tilde{f})^{(2)}, \nonumber \\ 
&&\hat{Q}_3=sf^{\dagger} + 
\chi_3(\tilde{d}f^{\dagger})^{(3)} + h.c., 
\label{op}
\end{eqnarray}

The model parameters are determined as follows.  The standard prescription 
for boson number (i.e.  counting pairs of nucleons above or below the 
nearest shell closure) would give $N=6$.  However, it is well known that 
the effective boson numbers are much smaller due to the $Z=64$ subshell 
closure \cite{CW88}.  The suggested effective numbers in the literature 
vary between $N=1$ and 3.  We adopted $N=2$ in our calculations, since 
there are experimental signatures for the two-phonon states, but no 
evidence for three-phonon states in $^{144}$Sm.  The parameters of the IBM 
Hamiltonian Eq.~(\ref{hibmsdf}) are simply determined from the excitation 
energies of the first $2^+$ and $3^-$ states in $^{144}$Sm as 
$\epsilon_d=1.66$ MeV and $\epsilon_f=1.81$ MeV. The nuclear potential 
parameters are taken from the exhaustive 
study of this reaction in 
Ref.~\cite{M95} as $V_0 = 105.1$ MeV, $R_0 = 8.54$ fm and $a = 0.75$ fm.  
Finally, the target radius is taken to be $R_T = 5.56$ fm.

The results of the coupled-channels calculations are compared with the 
experimental data in Fig.~4.  The upper and the lower panels in Fig.~4 show 
the excitation function of the fusion cross section and the fusion barrier 
distributions, respectively.  The experimental data are taken from 
Ref.~\cite{LDH95}.  The dotted line is the result in the harmonic limit, 
where couplings to the quadrupole and octupole vibrations in $^{144}$Sm are 
truncated at the single-phonon levels.  The deformation parameters are 
estimated to be $\beta_2$=0.11 and $\beta_3$=0.21 from the electric 
transition probabilities.  The dotted line reproduces the experimental data 
of both the fusion cross section and the fusion barrier distribution 
reasonably well, though the peak position of the fusion barrier 
distribution around $E_{cm}=65$ MeV is slightly shifted.  As was shown in 
Ref.~\cite{M95}, the shape of the fusion barrier distribution becomes 
inconsistent with the experimental data when the double-phonon channels are 
included in the harmonic limit (the dashed line).  The good agreement is 
recovered when one takes the effects of anharmonicity of the vibrational 
motion into account.  These results are shown in Fig.~4 by the solid line.  
This calculation has been performed using the parameters, $\beta_2=0.13$, 
$\beta_3=0.23$, $\chi_2=-3.30$, $\chi_{2f}=-$2.48, and $\chi_3$=2.87, which 
are obtained from a $\chi^2$ fit to the fusion cross sections.  The 
$\chi^2$ fit gave a unique result, regardless of the starting values.  The 
non-zero $\chi$ values indicate the anharmonic effects in the transition 
operators.  The slight change in the values of the deformation parameters 
from those in the harmonic limit results from the renormalization effects 
due to the extra terms in the operators given in Eq. (\ref{op}).  
Note that the 
solid line agrees with the experimental data much better than the dotted 
line.

One of the pronounced features of an anharmonic vibrator is that the 
excited states have non-zero quadrupole moments.  Using the 
$\chi$ parameters in the E2 operator, $T$(E2$)=e_B \hat Q_2$, 
extracted from the analysis of fusion 
data, we can estimate the static quadrupole 
moments of various states in $^{144}$Sm.  Here, $e_B$ is the effective 
charge, which is determined from the experimental $B($E2$;0\to 2^+_1)$ 
value as $e_B = 0.16~eb$.  For the quadrupole moment of the first 2$^+$ 
and 3$^-$ states, 
we obtain $-$0.28 b and $-$0.70 b, respectively.
The negative sign of the quadrupole moment of the octupole-phonon state is 
consistent with that suggested from the neutron pick-up reactions on 
$^{145}$Sm \cite{KGE89}.

In the case of rotational coupling, fusion barrier distributions strongly 
depend on the sign of the quadrupole deformation parameter through the 
reorientation term.  Also, as mentioned in the introduction 
it has been reported that fusion barrier 
distributions are very sensitive to the sign of the hexadecapole 
deformation parameter \cite{LLW93}.  Similarly, it is likely that the shape 
of fusion barrier distributions changes significantly when one inverts the 
sign of the quadrupole moment in a spherical target if there exists 
a strong anharmonicity.  Fig.~5 shows the 
influence of the sign of the quadrupole moment of the excited states on the 
fusion cross section and the fusion barrier distribution.  The solid line 
is the same as in Fig.~4 and corresponds to the optimal choice for the 
signs of the quadrupole moments of the first 2$^+$ and 3$^-$ states.  The 
dotted and dashed lines are obtained by changing the sign of the $\chi_2$ 
and $\chi_{2f}$ parameters in Eq.~(\ref{op}), respectively, while the 
dot-dashed line is the result where the sign of both $\chi_2$ and 
$\chi_{2f}$ parameters are inverted.  The change of sign of $\chi_2$ and 
$\chi_{2f}$ is equivalent to taking the opposite sign for the quadrupole 
moment of the excited states.  Fig.~5 demonstrates that subbarrier fusion 
reactions are indeed sensitive to the sign of the quadrupole moment of 
excited states.  The experimental data are reproduced only when the correct 
sign of the quadrupole moment are used in the coupled-channels 
calculations.  Notice that the fusion excitation function is completely 
insensitive to the sign of the quadrupole moment of the first $2^+$ state, 
but strongly depends on that of the first $3^-$ state.  In contrast, the 
fusion barrier distribution can probe the signs of the quadrupole moments 
of both the first $2^+$ and $3^-$ states.  This study shows that the sign of 
quadrupole moments in spherical nuclei can be determined from subbarrier 
fusion reactions, especially through the barrier distribution.

\medskip

\begin{flushleft} 
{\bf 4. Summary} 
\end{flushleft} 

We discussed the effects of anharmonicity 
in nuclear surface vibrations on heavy-ion fusion reactions by using 
the interacting boson model. Our analyses clearly show that 
the fusion barrier distribution is very sensitive to the anharmonicity 
of the nuclear surface vibrations, and suggest that the puzzle 
raised by Jack {\it et al.} concerning the $^{16}$O + $^{144}$Sm fusion 
reactions can be 
solved by considering the anharmonicities of the octupole surface 
vibrations of $^{144}$Sm. We have also shown that one can determine  
nuclear shape by using this high sensitivity of the 
fusion barrier distribution to the anharmonicity through reorientation 
effetcs. 

\bigskip

The authors thank J.R. Leigh, M. Dasgupta, D.J. Hinde, and J.R. Bennett for 
useful discussions.  K.H. and N.T. also thank the Australian National 
University for its hospitality and for partial support for this project.  
The work of K.H. was supported by the Japan Society for the Promotion of 
Science for Young Scientists.  This work was supported by the Grant-in-Aid 
for General Scientific Research, Contract No.06640368 and No.08640380, and 
the Grant-in-Aid for Scientific Research on Priority Areas, Contract 
No.05243102 and 08240204 from the Japanese Ministry of Education, Science 
and Culture, and a bilateral program of JSPS between Japan and Australia.

\newpage

\begin{center}
{\bf Figure Captions}
\end{center}

\noindent {\bf Fig.~1:} Effects of anharmonicity in the 
$^{16}$O + $^{154}$Dy fusion reactions. 

\noindent {\bf Fig.~2:} Dependence of the fusion barrier 
distribution on the energy spectrum and the boson number 
in the $^{16}$O + $^{154}$Dy fusion reactions. 

\noindent {\bf Fig.~3:} Dependence of the excitation function of the 
fusion cross section for the $^{16}$O + $^{154}$Dy reactions 
and of the fusion barrier distribution on the boson number.

\noindent {\bf Fig.~4:} Comparison of the experimental fusion cross section 
(the upper panel) and fusion barrier distribution (the lower panel) with 
the coupled-channels calculations for $^{16}$O + $^{144}$Sm reaction.  The 
experimental data are taken from Ref.~\cite{LDH95}.  The solid line shows 
the results of the present IBM model including the double-phonon states and 
anharmonic effects.  The dotted and the dashed lines are the results of the 
single- and the double-phonon couplings in the harmonic limit, 
respectively.

\noindent
{\bf Fig.~5:}
Dependence of the fusion cross section and barrier distribution on the 
sign of the quadrupole moment of the excited states in $^{144}$Sm.  The 
meaning of each line is indicated in the inset.

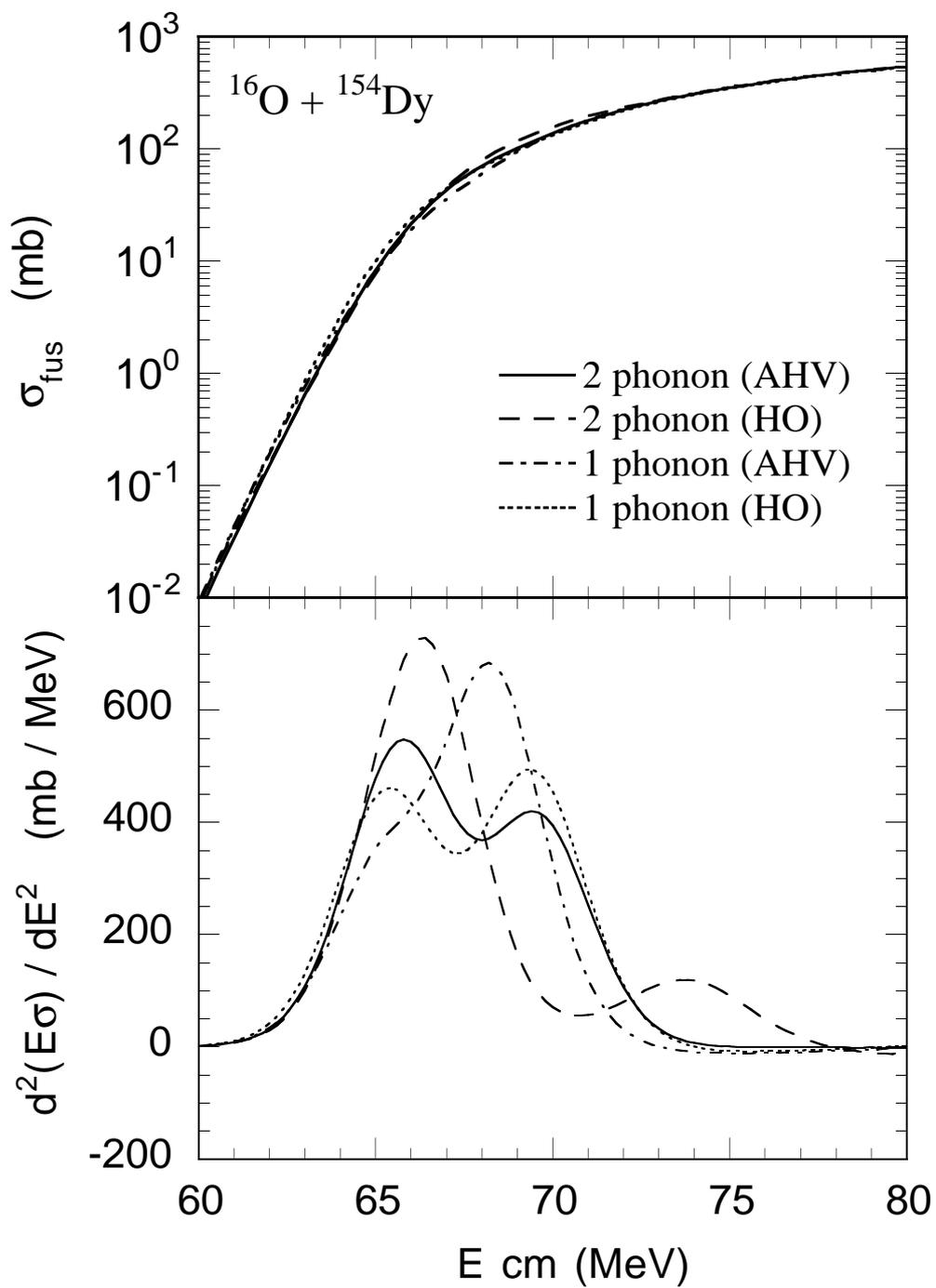
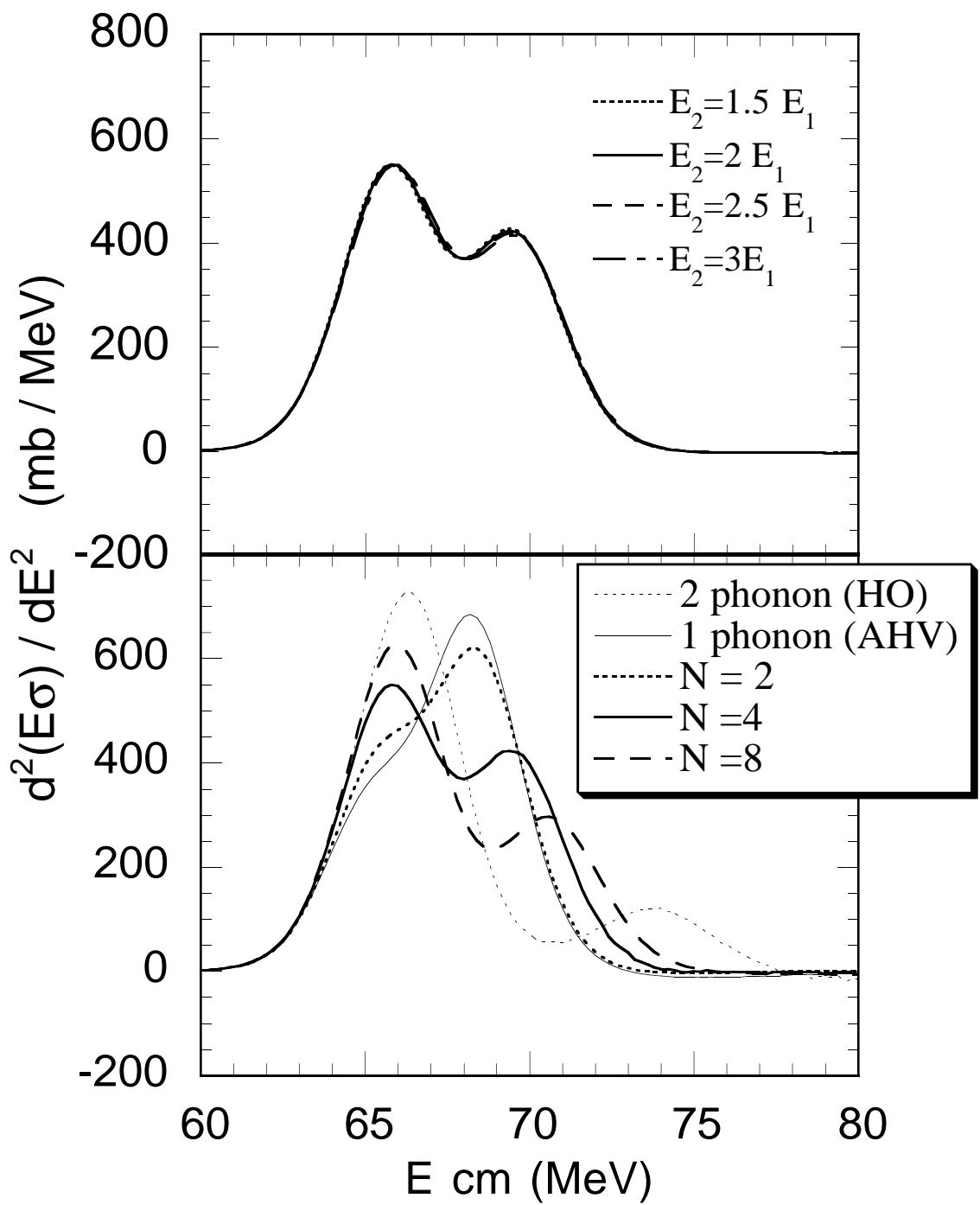
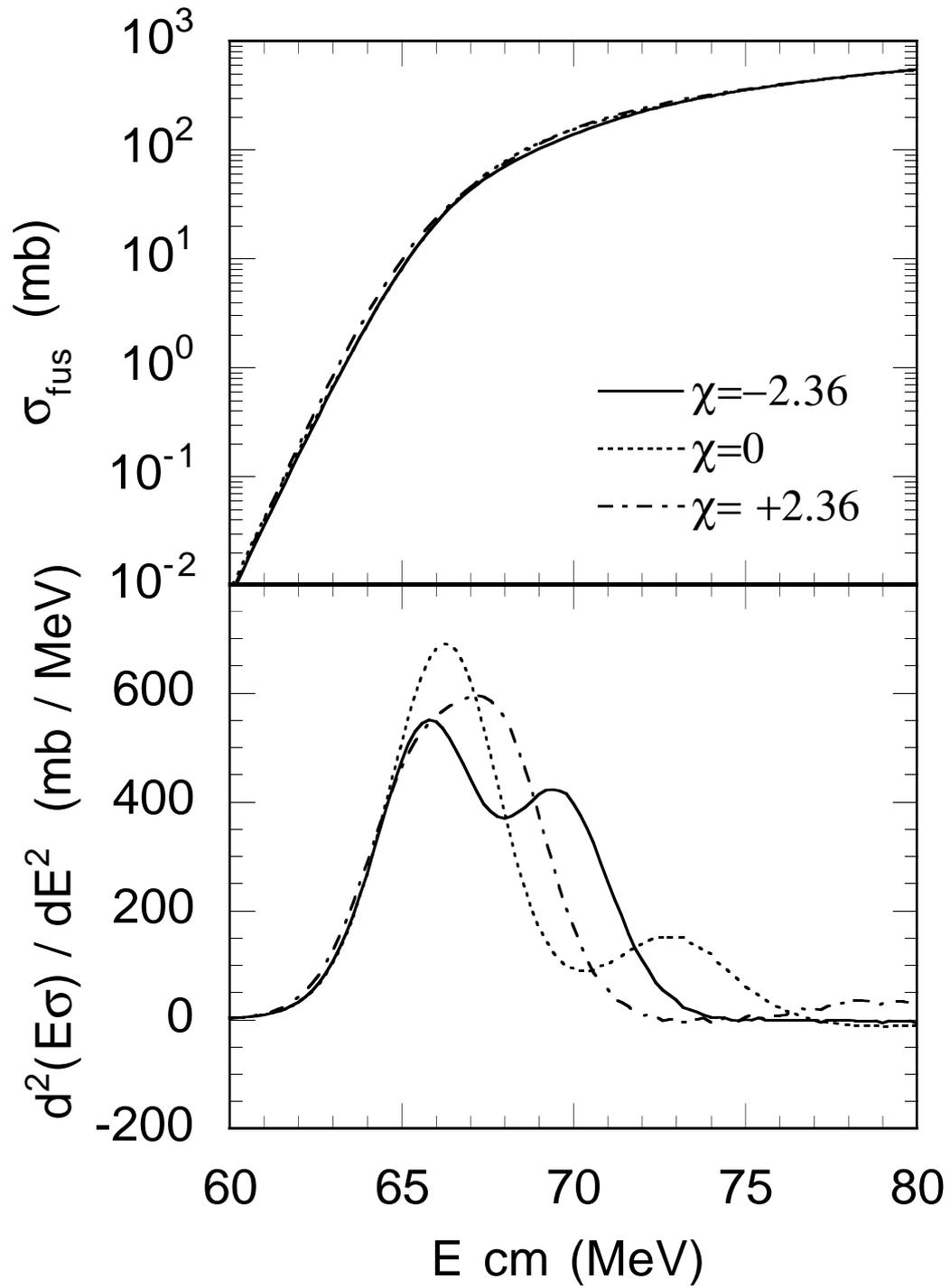
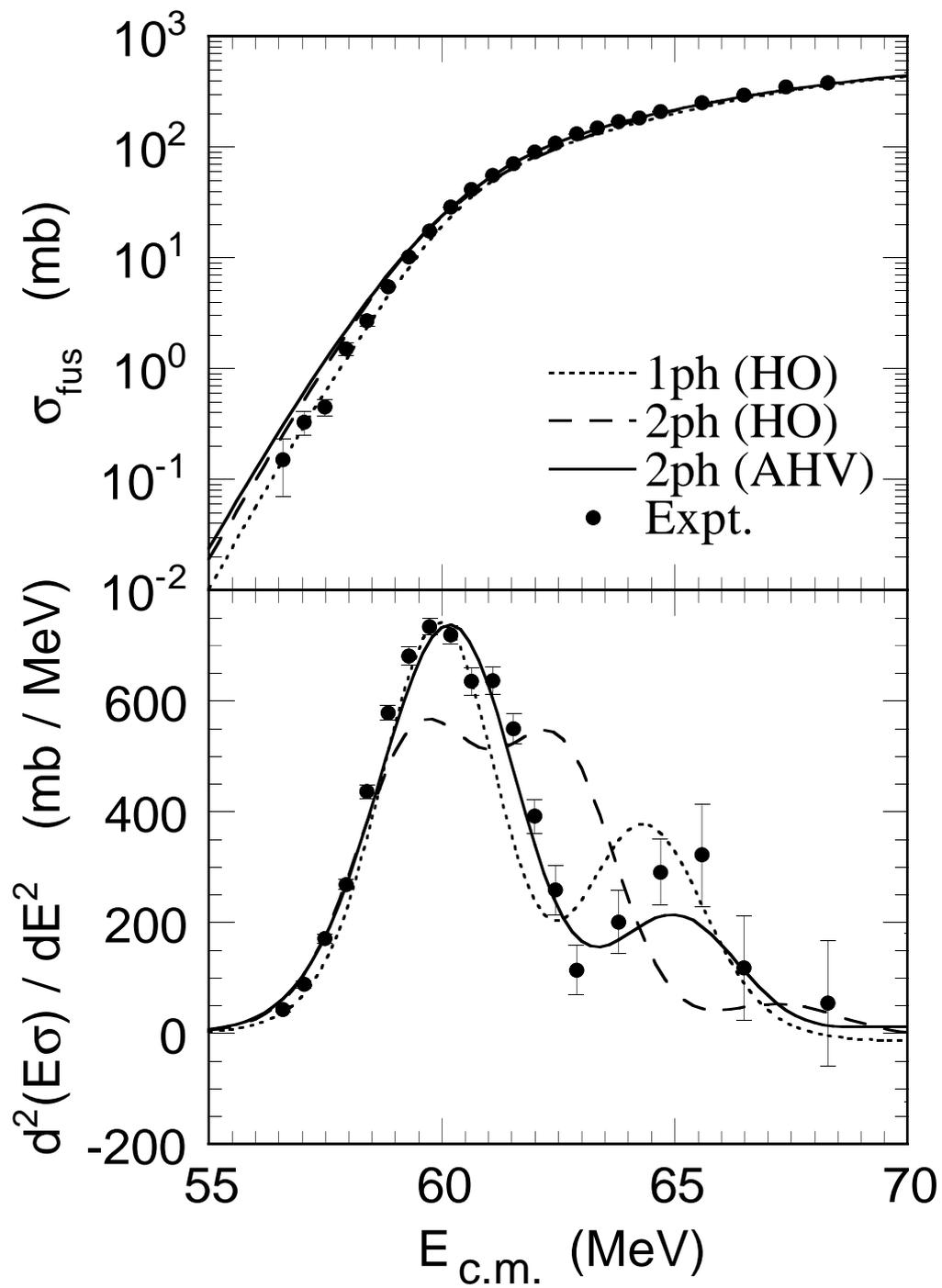
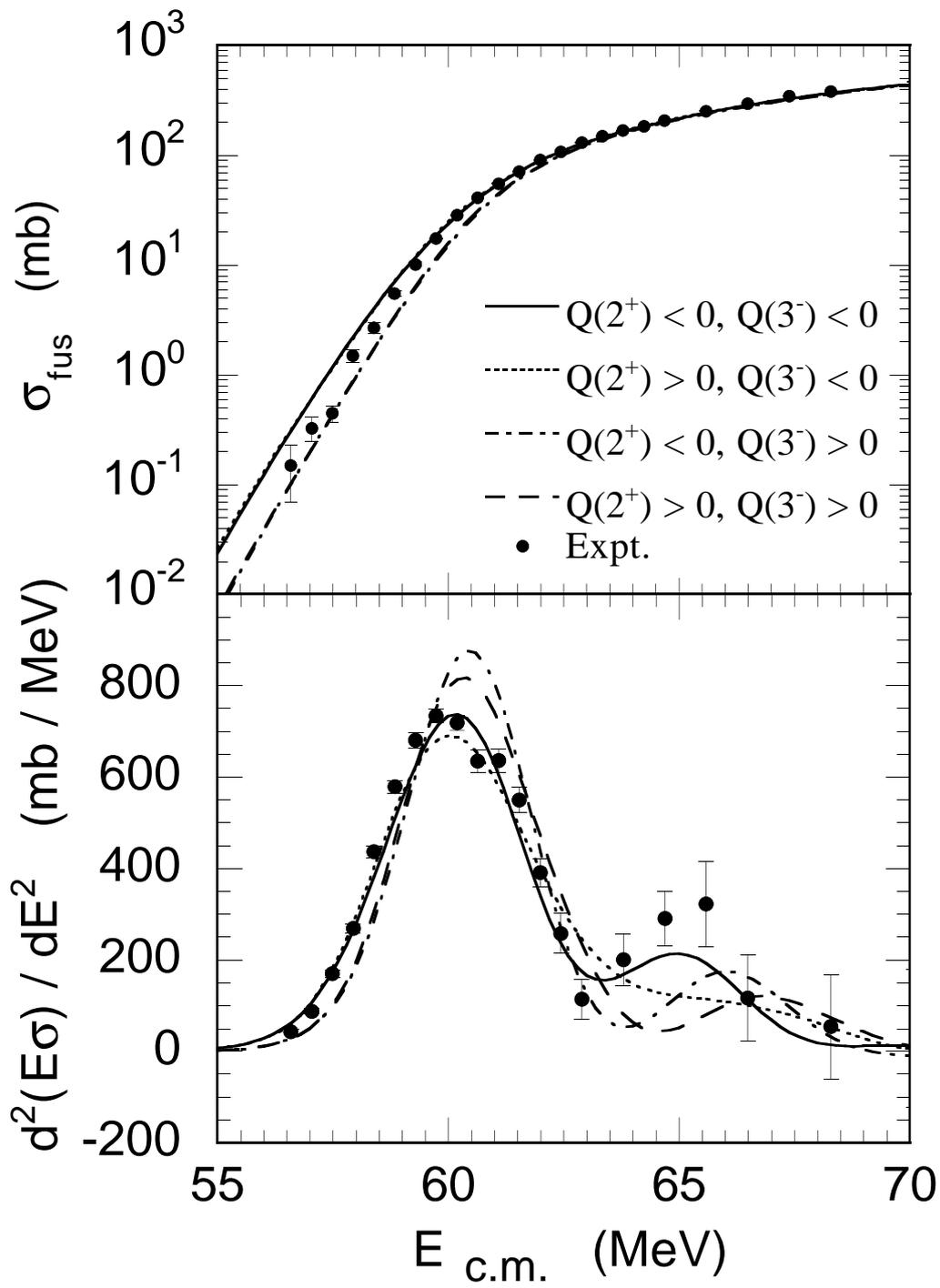
\end{document}